\documentclass[prl,twocolumn,showpacs,superscriptaddress,amsmath,amssymb]{revtex4-1} 
\usepackage{amsmath}
\usepackage{graphicx}
\usepackage{color}
\usepackage{amssymb}
\usepackage{ulem}
\usepackage[colorlinks=true,linkcolor=blue,citecolor=blue,urlcolor=blue]{hyperref}
\newcommand{\bea}{\begin{eqnarray}}
\newcommand{\eea}{\end{eqnarray}}
\newcommand{\be}{\begin{equation}}
\newcommand{\ee}{\end{equation}}

\newcommand{\rmS}{\text{S}}

\newcommand{\rmG}{\text{G}}
\newcommand{\rmE}{\text{E}}
\newcommand{\rmC}{\text{C}}
\newcommand{\rmB}{\text{B}}
\newcommand{\off}{\text{OFF}}
\newcommand{\on}{\text{ON}}
\newcommand{\rme}{\text{e}}

\newcommand{\kBT}{k_\text{B}T}

\definecolor{light-gray}{gray}{0.9}
\definecolor{gris}{gray}{0.5}

\begin{document}

\title{
\vspace{-2.25cm}
\textnormal{\small  PHYS. REV. B {\bf 95}, 241401(R) (2017)}\\
%\vspace*{-0.2cm}
%\rule[0.1cm]{18cm}{0.02cm}\\
%\ \\
\vspace*{0.285cm}
All-thermal transistor based on stochastic switching}
\author{Rafael S\'anchez}
\affiliation{Instituto Gregorio Mill\'an, Universidad Carlos III de Madrid, 28911 Legan\'es, Madrid, Spain}
\author{Holger Thierschmann}
\affiliation{Kavli Institute of Nanoscience, Faculty of Applied Sciences, Delft University of Technology, Lorentzweg 1, 2628 CJ Delft, The Netherlands}
\author{Laurens W. Molenkamp}
\affiliation{Experimentelle Physik 3, Physikalisches Institut, Universit\"at W\"urzburg, Am Hubland, 97074 W\"urzburg, Germany}
%\date{\today}

\begin{abstract}
Fluctuations are strong in mesoscopic systems and have to be taken into account for the description of transport. We show that they can even be used as a resource for the operation of a system as a device. We use the physics of single-electron tunneling to propose a bipartite device working as a thermal transistor. Charge and heat currents in a two-terminal conductor can be gated by thermal fluctuations from a third terminal to which it is capacitively coupled. The gate system can act as a switch that injects neither charge nor energy into the conductor, hence achieving huge amplification factors. Nonthermal properties of the tunneling electrons can be exploited to operate the device with no energy consumption.
\end{abstract}
%\pacs{
%73.23.-b, %Electronic transport in mesoscopic systems
%85.80.Fi, %Thermoelectric devices
%05.60.Gg %Quantum transport
%05.70.Ln	%Nonequilibrium and irreversible thermodynamics (Thermodynamics)
%85.80.-b	%Thermoelectromagnetic and other devices
%}
%\preprint: --------A-------- 
\maketitle

{\it Introduction.} %Understanding the properties of heat flows at small length scales is a key issue in nowadays electronics. 
Controlling the heat flow at small length scales is a great challenge in present-day electronics and a serious technological issue. 
%\sout{As the speed of operation, miniaturization and packing density further increases the dissipation of heat in micro electronic circuits becomes a serious issue.} 
On a scientific level, a significant effort is devoted to designing new concepts for nano scale heat engines and thermoelectric devices~\cite{holger,roche_harvesting_2015,koski_on-chip_2015,comptes} as well as finding new means to manipulate the flow of heat with devices such as thermal diodes and thermal transistors. This would not only allow for a more efficient removal of waste heat~\cite{giazotto_opportunities_2006}, but might also lead to the design of logic circuits operating with heat~\cite{li_colloquium_2012} or noise~\cite{pfeffer_logical_2015}. In recent years a variety of nano structures has been identified and demonstrated in experiments as possible thermal diodes~\cite{Chang_SolidState_2006,Scheibner_Quantum_2008,ruokola_single_2011,Martinez-Perez_Rectification_2015,diode}.
%\sout{due to their ability to rectify electronic or phononic heat currents}, 
%such as hybrid N-I-S junctions~\cite{Martinez-Perez_Rectification_2015}, quantum dots \cite{Scheibner_Quantum_2008,ruokola} or carbon nanotubes~\cite{Chang_SolidState_2006}. 
For a thermal transistor, however, the situation is more complicated because one has to efficiently switch heat currents by means of temperature. 
%In electronic systems, 
%\sout{which are the subject of this paper, }
%the main effect of temperature is known to be voltage fluctuations. Since the Fermi distribution is symmetric with respect to energy, the problem arises how to switch a thermal conductor between conductive and insulating states with an energy-symmetric parameter.
%In this letter we introduce a new approach for a thermal transistor which is based on stochastic switching of heat currents that are carried by particles in a solid state device. We show that such a transistor can in principle operate as an ideal device with a divergin amplification factor. We note that the mechanism  has been demonstrated recently in experiments, indicating its relevance for pratcial application.

\begin{figure}[b]
%\begin{center}
%\includegraphics[width=\linewidth,clip] {./figs/dqdscheme.eps}
\includegraphics[width=0.9\linewidth,clip] {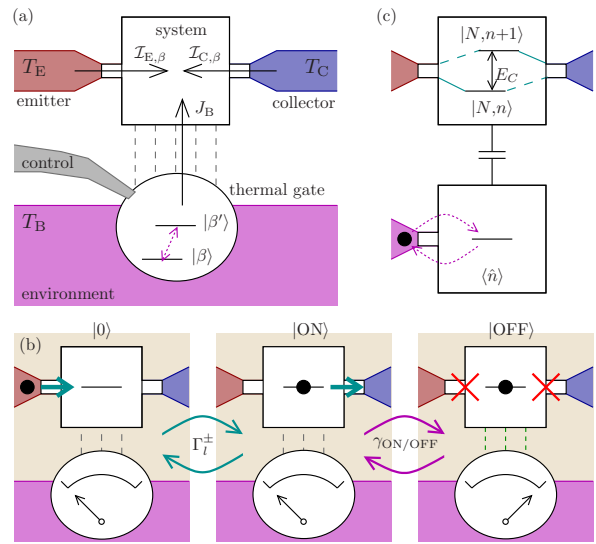}
%\end{center}
\caption{
\label{scheme} 
%All-thermal transistor based on thermal gating. 
(a) A mesoscopic thermal gate mediates the system-environment interaction. Currents ${\cal I}_{l,\beta}$ in the emitter and collector terminals are affected by the thermal gate whose state $|\beta\rangle$ is manipulated by changing the temperature $T_\rmB{\to}T_\rmB{+}\Delta T_\rmB$. (b) A minimal three-state model treating the gate as a switch achieves the transport modulation with no heat exchange. (c) Two sites with a strong Coulomb interaction. The occupation of the gate, $n$, affects transport through the system.
}
\end{figure}

Recent proposals suggest the use of nonlinearities of a mesoscopic system coupled to environmental modes~\cite{li_negative_2006,segal_nonlinear_2008,ruokola_thermal_2009,jiang_phonon_2015,
benenti_from_2016,joulain_quantum_2016}.
Usually, a system is considered that is connected to two terminals, emitter E and collector C, and an environment, acting as a base B. A temperature distribution $\{T_k\}=(T_\rmE,T_\rmC,T_\rmB)$, generates transport in the system. The aim is to modulate the collector heat current $J_\rmC$ with a small modulation of the heat injected from the base, $J_\rmB$. 
This is usually done via inelastic transitions in the system induced by fluctuations in the environment. These can be controlled by tuning the base temperature $T_\rmB{\to}T_\rmB{+}\Delta T_\rmB$.
%These are at temperatures $T_\rmE$ and $T_\rmC$, respectively.
%Then, fluctuations in the environment B induce inelastic transitions in the system which affect the transport through the system. 
%They also involve an energy exchange $J_\rmB$. Thereby the heat currents $J_l(\Delta T_k)$, induced by a temperature gradient $\Delta T_k$, with $k,l=\rmE,\rmC$, become modulated by a change in the temperature of the environment, $\Delta T_\rmB$. 
A thermal transistor effect appears whose amplification factor is defined as~\cite{li_negative_2006,jiang_phonon_2015,joulain_quantum_2016}:
\be
\label{amplfact}
\alpha_{l}=\frac{\Delta J_l(\Delta T_\rmB)}{\Delta J_\rmB(\Delta T_\rmB)},
\ee
with $l$=E,C and $\Delta J_k(\Delta T_\rmB){=}J_k(\Delta T_\rmB){-}J_k(0)$. The challenge is now to have a sizable modulation $\Delta J_l$ out of tiny injected currents, $\Delta J_\rmB$. However, energy exchange with the environment is inherent to inelastic transitions, thus limiting the performance of the transistor. 
%Since $\Delta J_\rmB$ is determined by the energy exchange between the environment and the system, this interaction is the crucial parameter.
%In general, the state of the system is affected by the presence of an environment with which it interacts. Usually this interaction induces or affects rates of state transitions of the system. The environment-assisted rates then depend on the temperature of the environment. If the system serves as a link between two reservoirs, transport will depend on its state and thus on the temperature of the environment. However, these transitions being inelastic involve energy exchange that limits the transistor performance.

Here we propose an alternative approach. We introduce a nanostructure that mediates the coupling between the system and the environment, which we call a \textit{thermal gate} [cf. Fig.~\ref{scheme}(a)]. Transport in the system then becomes dependent on the state of the gate, which fluctuates at a rate given by the temperature $T_\rmB$. In particular, the gate can act as a switch if a given (thermally-activated) transition blocks the system currents, with no net energy exchange involved. This way, huge amplification coefficients can be achieved.

An additional advantage of our mechanism is that the state of the gate (and hence how the system is coupled to the environment) can be externally controlled at a microscopic level. This scheme furthermore helps to isolate the system from undesired (e.g. phononic) degrees of freedom. The system-gate interaction can be of a different nature, depending on the particular configuration. As an example, the Coulomb interaction has been used in the last years to investigate effects such as mesoscopic Coulomb drag~\cite{drag,bischoff_measurement_2015,keller_cotunneling_2016}, energy harvesters~\cite{hotspots,holger}, Maxwell demon refrigerators~\cite{koski_on-chip_2015}, or heat engines with no heat absorption~\cite{whitney_thermoelectricity_2016}. Most remarkably for us, stochastic switching of transport due to coupling to a mesoscopic gate has been recently demonstrated~\cite{singh_distribution_2016}.

%tuning the gate controls how the system is coupled to environmental fluctuations at a microscopic level, furthermore helping to isolate the system from undesired (e.g. phononic) degrees of freedom. In particular, the system is sensitive to fluctuations in the state of the thermal gate only, which thus acts as a switch on currents. Recently effects based on this mechanism, such as mesoscopic Coulomb drag~\cite{drag,bischoff_measurement_2015,keller_cotunneling_2016}, energy harvesters~\cite{hotspots,holger}, Maxwell demon refrigerators~\cite{koski_on-chip_2015}, or heat engines with no heat absorption~\cite{whitney_thermoelectricity_2016} have been investigated. Most remarkably for us, stochastic switching of transport due to coupling to a mesoscopic gate has been demonstrated~\cite{singh_distribution_2016}. We will show that it can be used to design a highly efficient all-thermal transistor  for which the effect on heat transport can be maximized at a minimal, and even zero, energy injected from the base environment.

%~\cite{drag,hotspots,holger,roche_harvesting_2015,hartmann_voltage_2015,pfeffer_logical_2015,thgating}
%Based on this principle it has been shown that charge currents can be manipulated by means of temperature for instance in superconducting weak links \cite{Morpurgo_Hot_1998} and quantum dot systems \cite{thgating}. 
%%%%%%%%%%%%%%%%%%%%%%%%%%%%%%
%%%%%%%%%%%%%%%%%%%%%%%%%%%%

{\it Three-state case.} Let us first illustrate the idea with a simple toy model.  
We assume that the device can exhibit three states which we label $|m\rangle=|0\rangle$, $|\on\rangle$, and $|\off\rangle$, cf. Fig.~\ref{scheme}(b). Transport occurs via transitions $|0\rangle\leftrightarrow|\on\rangle$ through leads $l{=}\rmE,\rmC$ at a rate $\Gamma_l^\pm$. The superindex $+(-)$ accounts for transitions populating (depopulating) the system. The $|\off\rangle$ state is uncoupled from transport and detuned by $\Delta E$. Stochastic (Markovian) transitions of the type $|\on\rangle{\leftrightarrow}|\off\rangle$ switch the system currents. These are due to fluctuations in the gate and are hence assisted by the environment with rates $\gamma_{\rm ON}$ and $\gamma_{\rm OFF}=\rme^{-\Delta E/\kBT_B}\gamma_\on$ obeying detailed balance. 

%The $|\off\rangle$ state is uncoupled from transport and detuned by $\Delta E$. Hence, there are only two possible types of transitions: $|0\rangle\leftrightarrow|\on\rangle$ (conducting), and $|\on\rangle\leftrightarrow|\off\rangle$ (switching), cf. Fig.~\ref{scheme}(b). The conducting transitions through contact $l{=}\rmE,\rmC$ are given by the transport rates $\Gamma_l^\pm$, where the superindex $+(-)$ indicates transitions populating (depopulating) the system. These rates are given by the details of coupling between system and its leads. In contrast, the rates for switching, $\gamma_{\rm ON}$ and $\gamma_{\rm OFF}$, are due to fluctuations in the gate and are therefore assisted by the environment. 

We can write a rate equation for the probability of the system states $P_m$:
\begin{align}
\dot P_0&=\Gamma_\Sigma^-P_\on-\Gamma_\Sigma^+P_0\\
\dot P_\off&=\gamma_{\off}P_\on-\gamma_{\on}P_\off,
\end{align}
with $1=\sum_m P_m$ and $\Gamma_{\Sigma}^\pm=\sum_l\Gamma_l^\pm$.
Solving for the stationary state, $\dot P_m=0$, we can write the particle currents, ${\cal I}_\rmC=-{\cal I}_\rmE=\Gamma_\rmC^+P_0-\Gamma_\rmC^-P_\on$, giving
\be
\label{eq:curr3st}
{\cal I}_\rmC={\cal I}_\rmC^{\on}\left(1-P_\off\right).
\ee
It conditions the ``uncoupled" current ${\cal I}_\rmC^{\on}$ on the depopulation of the OFF state, with
\be
P_\off(T_\rmB)=\frac{\rme^{-\Delta E/\kBT_\rmB}\Gamma_\Sigma^+}{\Gamma_\Sigma^-+\Gamma_\Sigma^+\left(1+\rme^{-\Delta E/\kBT_\rmB}\right)}.
\ee
Remarkably, it does not depend on the details of the coupling to the environment, only on its temperature. 

In this configuration, the particle current (regardless of the electric or thermal gradient that originates it) is directly proportional to the charge and heat currents $I_l=e{\cal I}_l$ and $J_l=(E_\on-\mu_l){\cal I}_l$, with $e$ being the elementary charge, $E_\on$ the energy of the state and $\mu_l$ the electro-chemical potential of $l$.
This way, any current in the conductor can be modulated by tuning $T_\rmB$, in particular: 
\be
%\Delta J_l(\Delta T_k)=J_l(\Delta T_k,0)\Delta P_\off(\Delta T_\rmB),
\Delta J_\rmC=-J_\rmC^\on\Delta P_\off,
%\frac{P_\off(0)-P_\off(\Delta T_\rmB)}{1-P_\off(0)},
%\frac{\sum[\Gamma_l^++\Gamma_l^-(1-\rme^{-\Delta/kT_\rmB})]}{\sum[\Gamma_l^-+\Gamma_l^+(1+\rme^{-\Delta/kT_\rmB})]},
\ee
with $\Delta P_\off{=}P_\off(T_\rmB{+}\Delta T_\rmB){-}P_\off(T_\rmB)$. Remarkably this modulation takes place without any energy exchange with the environment. Therefore, the amplification factor $\alpha_{l}$ in Eq.~\eqref{amplfact} diverges. This makes the device an ideal all-thermal transistor. We note that it could also be used as a perfect and non-invasive thermometer: Changes in the temperature of the environment are measured in the modulation of a (electrically or thermally generated) charge current: $\Delta I_l={-}I_l^\on\Delta P_\off$.

%%%%%%%%%
%%%%%%%%%
{\it Implementations.} Let us now specify a possible physical realization of the system/thermal gate partition. After presenting a description of the heat currents we show how the model can be mapped on the ideal three-state case described above in the appropriate configuration.

The minimal model for transport consists of a single site that connects the two terminals $l$=E,C. This site can be empty or occupied. The thermal gate consists of a similar site which is coupled to reservoir B. Both subsystems are allowed to fluctuate between two states due to the coupling to three different terminals [Fig.~\ref{scheme}(c)].

\begin{figure}[t]
%\begin{center}
\includegraphics[width=\linewidth,clip] {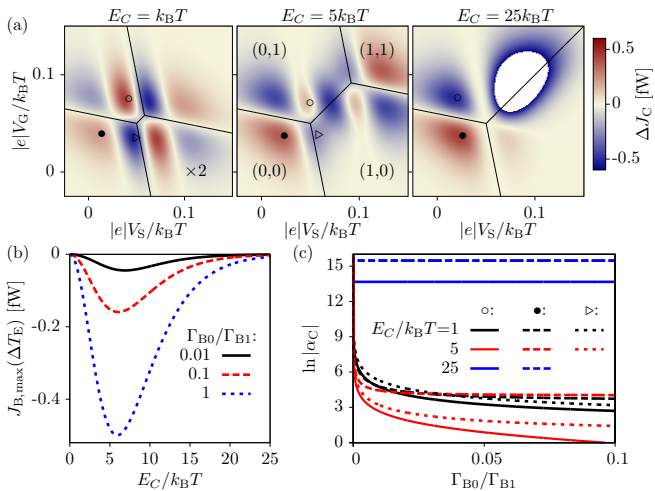}
%\end{center}
\caption{\label{thgatEc}(a) Thermal gating $\Delta J_\rmC$ as a function of the partition energies for different values of the interaction $E_C$ in the case $\Gamma_{ln}=10\Gamma_{\rmB N}=10$~$\mu$eV$/\hbar$. Solid lines mark the degeneracies of the different occupation probabilities $P_{(N,n)}$ which cross at triple points. A negative contribution in the rightmost panel is cut off for clarity. Even if the gating is larger in that region, the amplification is smaller for being closer to the maximum of $J_\rmB$. (b) Maximal heat exchange $J_\rmB$ as a function of the interaction $E_C$. Energy-dependent tunneling in the gate reduces the heat current. (c) Amplification factor as a function of the  asymmetry $\Gamma_{\rmB0}/\Gamma_{\rmB1}$ for the points with enhanced $|\alpha_{\rmC}|$ labeled in (a). Parameters: $T_{\rmC,\rmB}=T=0.243$~K, $T_{\rmE}=3T/2$, $\Delta T_\rmB=T/2$, $\alpha_i=0.1$, $\beta_i=0.002$.
}
\end{figure}

In ultrasmall devices with strong Coulomb interactions, each site can be assumed to be occupied by up to one electron. Four states $|N,n\rangle$ are thus relevant, with $N,n=0,1$ being the charge of the system and of the thermal gate. For simplicity, we consider single-level systems as can be found, for example, in semiconductor quantum dots~\cite{bischoff_measurement_2015,holger,thgating}. We remark, however, that our main results are general and also apply, for instance, to cavities with a high density of states~\cite{hartmann_voltage_2015,pfeffer_logical_2015,roche_harvesting_2015} or to metallic single-electron transistors~\cite{koski_on-chip_2015,singh_distribution_2016}.
The Hamiltonian of the bipartite system is given by 
%$\hat H=\hat H_{\rmS}+\hat H_{\rmG}+\hat H_{\rmS\rmG}$, with
\be
%\hat H_i=\varepsilon_i\hat c_i^\dagger \hat c_i+\sum_{lk}\hat \varepsilon_{lk}\hat d_{lk}^\dagger \hat d_{lk}+\sum_{l,k} (v_{lk}\hat c_{i}^\dagger d_{lk}+{\rm h.c.}),
\hat H=\varepsilon_\rmS\hat N+\varepsilon_\rmG\hat n+E_C\hat N\hat n,
\ee
%for $i$=S,G labeling the system and the thermal gate, respectively. The operators $c_i$ and $d_{lk}$ anhilate and electron in dot $i$ and terminal $l$ (with momentum $k$), respectively. These two subsystems are coupled via $\hat H_{\rmS\rmG}=E_C\hat N\hat n$, 
with $\hat N$ and $\hat n$ being the number operators in the sytem and gate dots. The Coulomb interaction shifts the energy of a site by $E_C$ when the respective other site is charged~\cite{hotspots}. This term introduces the correlation of the system currents with the state of the thermal gate. The device charging can be controlled experimentally by means of gate voltages, $V_\rmS$ and $V_\rmG$ [cf. Fig.~\ref{thgatEc}(a)]. Then the energy of the respective site is $\varepsilon_{i}=\varepsilon_i^{(0)}+\alpha_ieV_i+\beta_ieV_{j}(1-\delta_{ij})$, where $\varepsilon_i^{(0)}$ is the on-site bare energy, and $\alpha_i$, $\beta_i$ are constants given by cross-capacitances~\cite{hotspots}.

We assume the weak coupling limit, $\hbar\Gamma_{lq}\ll\kBT_l$, where transport is well described by sequential tunneling rates which are quite generally energy dependent: $\Gamma_{lq}^\pm=\Gamma_{lq}f_l^\pm(U_{iq})$. Here $U_{iq}=\varepsilon_i+E_C\delta_{q1}$, $q$ is the occupation of the respective other subsystem, $f_l^+(E)=[1+\rme^{(E-\mu_l)/\kBT_l}]^{-1}$ is the Fermi function, and $f_l^-(E)=1-f_l^+(E)$~\cite{drag}. For simplicity, we write $\Gamma_l(U_{lq})=\Gamma_{lq}$. We are mainly interested in configurations close to at least one triple point (with $|U_{iq}|<\kBT_l$ for every contact) where the current is enabled by charge fluctuations in both dots (cf. Fig.~\ref{thgatEc}). There, charging and uncharging rates are of the same order in all barriers, $\Gamma_{lq}^+\sim\Gamma_{lq}^-$. Higher-order tunneling effects~\cite{cotunneling} can be neglected in this limit.  

We write four rate equations for the occupation of the different states, $P_{(N,n)}$~\cite{beenakker,hotspots}, whose stationary solution gives the
state-resolved particle currents~\cite{detection},
\be
\label{strescurr}
{\cal I}_{lq}=\Gamma_{lq}^+P_{(0,q)}-\Gamma_{lq}^-P_{(1,q)},\text{ for $l$=E,C}.
\ee
For ${\cal I}_{\rmG q}$, replace $P_{(i,q)}{\to}P_{(q,i)}$.
From them we obtain the charge $I_l=e\sum_q{\cal I}_{lq}$ and heat currents $J_l=\sum_q(U_{lq}{-}\mu_l){\cal I}_{lq}$. We remark that both the energy and rates of each subsystem depend on the charge state of the other one.

In such a device, cyclic transitions exist which transfer an energy $E_C$ between the system and the gate \cite{drag,holger,bischoff_measurement_2015,hartmann_voltage_2015}. They are detrimental for our purpose of gating the system with minimal heat exchange. These transitions are of the form $|0,0\rangle\leftrightarrow|1,0\rangle\leftrightarrow|1,1\rangle\leftrightarrow|0,1\rangle\leftrightarrow|0,0\rangle$ including all four charge states. Hence, energy transfer can be suppressed by selecting configurations where transitions in one of the systems are conditioned on the state of the other one, i.e. if $\Gamma_{ln}^\pm=0$ for all terminals $l$ in one partition with the other one being in state $n$. This effectively reduces the system to the three-state case discussed above. We have found two ways how this can be achieved: (i) by filtering some of the transitions in one of the partitions by highly energy selective tunneling or (ii) by increasing the interaction energy $E_C$.

Let us first consider case (i) with, e.g. $\Gamma_{\rmB0}{=}0$ such that transitions $|0,0\rangle\leftrightarrow|0,1\rangle$ are avoided. 
%Only transitions $|0,0\rangle\leftrightarrow|1,0\rangle\leftrightarrow|1,1\rangle$ are hence possible. 
This extremely energy-dependent tunneling can be achieved, e.g. by introducing a resonance (a second quantum dot) or a gap in the contact with the base. Then state-resolved currents are conserved: It is clear that ${\cal I}_{\rmB0}=0$ and (from charge conservation) also ${\cal I}_{\rmB1}=0$ and ${\cal I}_{\rmE n}=-{\cal I}_{\rmC n}$~\cite{detection}. Therefore from Eq.~\eqref{strescurr} we have $J_\rmB=0$, up to higher order tunneling corrections~\cite{cotunneling}, neglected here. Note that filtered transitions furthermore suppress the eventual contribution of non-local cotunneling of the form $|1,0\rangle\leftrightarrow|0,1\rangle$. Despite the absence of an energy exchange, the state of the system is still sensitive to the occupation in the thermal gate. Thus the particle current reads
\be
{\cal I}_\rmC={\cal I}_\rmC^{\rm iso}(U_{\rmS 1})\langle \hat n\rangle+{\cal I}_\rmC^{\rm iso}(U_{\rmS 0})(1-\langle \hat n\rangle),
\ee
in terms of $\langle \hat n\rangle=P_{(0{,}1)}+P_{(1{,}1)}$ and the energy resolved current of an isolated conductor~\cite{buttiker_coherent_1988},
\be
{\cal I}_\rmC^{\rm iso}(E){=}\left[\frac{1}{\Gamma_{\rmE}(E)}{+}\frac{1}{\Gamma_{\rmC}(E)}\right]^{-1}{[f_\rmC ^+(E){-}f_{\rmE}^+(E)]}.
%\frac{\Gamma_{\rmE}(E){+}\Gamma_{\rmC}(E)}{\Gamma_{\rmE}(E)\Gamma_{\rmC}(E)}{[f_\rmC ^+(E){-}f_{\rmE}^+(E)]}.
\ee
The system switches between two currents depending on the state of the gate: ${\cal I}_l^{\rm iso}(U_{\rmS 1})$, when the gate is occupied, and ${\cal I}_l^{\rm iso}(U_{\rmS 0})$, when the gate is empty. Note that the thermodynamic state of the base reservoir, only enters through the average occupation of the thermal gate $\langle \hat n\rangle$. Hence we obtain a switching effect in the system which is solely driven by the fluctuations of the gate,
\be
\label{eq:thgatfilter}
\Delta {\cal I}_l{=}\left[{\cal I}_l^{\rm iso}(U_{\rmS 1}){-}{\cal I}_l^{\rm iso}(U_{\rmS 0})\right]\langle \hat n(\Delta T_\rmB){-}\hat n(0)\rangle.
\ee
We emphasize that switching takes place with only negligible leakage from the base terminal, thus leading to a large amplification factor.

%The amplitude and sign of $\Delta {\cal I}_l$ strongly depend on the potential of the mesoscopic regions, which shown in Fig.~\ref{thgatEc}(a) for the case of $J_\rmC$ as a function of $V_\rmS$ and $V_\rmG$ \textcolor{blue}{[is this true? If not we can also move this passage to the Ec section.]}. 
%For example, when $\Gamma_{\rmE0}=\Gamma_{\rmC0}=0$, transport is cancelled in the conductor when the gate is empty,~\cite{long} hence corresponding to the switch off state. The switched-on particle current in the collector ${\cal I}_{\rmC n}=(\Gamma_{\rmE n}^{-1}+\Gamma_{\rmC n}^{-1})^{-1}[f_\rmC ^+(\varepsilon_{\rmS n})-f_{\rmE}^+(\varepsilon_{\rmS n})]$ is modified by the fluctuations of the gate, so the dc current reads:
%\be
%{\cal I}_l={\cal I}_{ln}\langle n\rangle.
%\ee
%Note that the thermodynamic state of the base reservoir only enters through the average occupation of the switch. Hence the thermal gating effect is driven by the fluctuations of the gate system only. As we have $J_\rmB=0$, the amplification factor diverges. The previous result requires the control of the tunneling processes in two barriers. Note that similar results can be obtained in a less complicated configuration by filtering only the gate contact ($\Gamma_{\rmB N}=0$, for a particular $N$). 

The effect of thermal gating even improves for the second case (ii), where we control the interaction energy [see Fig.~\ref{thgatEc}(a)]. 
%As we will see this does not only suppress $J_B$ but it also increases the first term in eq.~\ref{eq:thgatfilter}, which describes the particle flow \textcolor{blue}{[not sure if this is the best way of putting it.]}.
%This  as we show in Fig.~\ref{thgatEc} by fixing the temperature and increasing $E_C$. 
Increasing $E_C$ can be done experimentally by electrostatic bridging of the two systems~\cite{chan_strongly_2002,hubel_two_2007} or in stacked two-dimensional materials, e.g., graphene~\cite{bischoff_measurement_2015}. 
We  start by considering the strong interaction limit, $E_C\gg \kBT$. We can then choose a configuration close to a triple point for which fluctuations occur in sequences of the form $|N{\pm}1{,}n\rangle{\leftrightarrow}|N{,}n\rangle{\leftrightarrow}|N{,}n{\pm1}\rangle$ with either $N{=}n{=}0$ (${+}$) or $N{=}n{=}1$ ($-$)~\cite{singh_distribution_2016}. Hence we have the seemingly paradoxical consequence that the transferred heat vanishes when the interaction of the two systems is large, $J_\rmB\rightarrow0$ for $E_C\rightarrow\infty$ [cf. Fig.~\ref{thgatEc}(b)]. 
%The full counting statistics of such a configuration has been recently reported in single-electron transistors.

Indeed, this limit can be mapped to the ideal three-state system discussed above, if $|\on\rangle=|N{,}n\rangle$, $|\off\rangle=|N{,}n{\pm}1\rangle$ and $|0\rangle=|N{\pm}1{,}n\rangle$, with $\Delta E=\pm(U_{\rmG n}-\mu_\rmB)$, and $n$ given by the charge occuparion of the $|\on\rangle$ state. The switching transitions are now given by $\Gamma_{\rmB N}^\pm$, which obviously satisfy local detailed balance. We therefore recover the solution obtained in Eq.~\eqref{eq:curr3st}. Then the linear thermal gating can be expressed as
\be
\label{linear}
\frac{1}{J_l}\frac{\partial J_l}{\partial T_\rmB}=\frac{U_{\rmG n}-\mu_\rmB}{\kBT_\rmB^2}\left(n-\langle\hat n\rangle\right),
\ee
for a small gradient $\Delta T_\rmB$ but arbitrary temperature configuration $\{T_k\}$.
%where $n=0$ for the case with $|\on\rangle=|0,0\rangle$, and $n=1$ for $|\on\rangle=|1,1\rangle$. Eq.~\eqref{linear} emphasizes the charge noise in the gate dot, $\langle(\Delta \hat n)^2\rangle$.

Figure \ref{thgatEc}(b) shows the two ways for minimizing the exchanged energy $J_\rmB$ discussed above: increasing $u=E_C/\kBT$ and introducing level-selected rates.
%where the device operates as a perfect thermal amplifier. 
An ideal behaviour is found at $u\gg1$, for strongly coupled systems or at low temperatures. For smaller $u$, undesired cyclic fluctuations involving all the four states start to contribute. These can however be filtered by preventing some tunneling transitions, e.g. by making $\Gamma_{\rmB0}\ll\Gamma_{\rmB1}$, as discussed above.
In Fig.~\ref{thgatEc}(c) we plot the interplay of these two mechanisms. As expected, the amplification factor increases for $\Gamma_{\rmB0}/\Gamma_{\rmB1}\rightarrow0$. For small but finite ratios we observe a non-monotonic behaviour as the interaction $u$ increases, with a larger amplification for small $u$. Note that it remains almost constant for large $u$, i.e. when the base current is suppressed. This is an indication that the thermal gating is purely induced by fluctuations that only depend on the temperature $T_\rmB$. Remarkably, the largest amplification is found in configurations ($\bullet$) close to one triple point, where sequential tunneling dominates even if one relaxes the weak-coupling assumption.

{\it Energy consumption.} Even if no heat flows through the contact to the base terminal, in order to operate the transistor one has to repeatedly inject (remove) a finite amount of heat in order to increase (reduce) its temperature. In this case the operation speed of the transistor is limited by the time scale of thermalization of the base terminal. 
%Having nanostructured contacts is also beneficial in this sense, as no temperature can be defined for the single-electron in the quantum dots~\cite{whitney_thermoelectricity_2016}. 
However, as thermal gating is driven by charge fluctuations, these can be experimentally frozen by simply closing the tunneling barrier with a plunger gate~\cite{holger} which does not only allow for much higher switching rates but also can be done at an arbitrarily low energy cost.
%
%{\it Energy consumption---} Even if no heat flows through the contact to the base terminal, one in principle has to inject a finite amount of heat in order to increase its temperature. And this has to be done repeatedly. Having nanostructured contacts is also beneficial in this sense, as no temperature can be defined for the single-electron in the quantum dots~\cite{whitney_thermoelectricity_2016}. The gating is driven by charge fluctuations and the rate for switching. These can be frozen by simply closing the tunneling barrier with a plunger gate, so one can fully control the thermal gating at will. Furthermore, opening/closing the gate costs arbitrarily low energy. 

\begin{figure}[t]
%\begin{center}
\includegraphics[width=0.8\linewidth,clip] {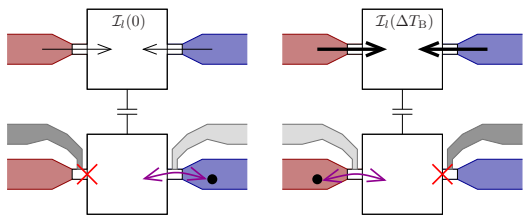}
%\end{center}
\caption{\label{operation} Operation of a thermal transistor at arbitrarily  low energy consumption. Opening/closing the tunneling barriers of the gate system coupled to two reservoirs (one hot, one cold), modulates the system currents without needing to let the base reservoir heat up or cool down each time.
}
\end{figure}
Alternatively, the gate system can be coupled to two reservoirs: one at temperature $T$ and the other one at $T+\Delta T_\rmB$, as depicted in Fig.~\ref{operation}. By opening or closing either of the two barriers, the fluctuations will be governed by either of the two temperatures. The two base reservoirs could even be in thermal contact with the emitter and the collector, so the current is driven and modulated by a single temperature gradient. In this case the speed of operation is limited by the time scale of charge relaxation rates. A periodical application of this mechanism would also enable a time-dependent driving of the heat currents in the conductor.

{\it Conclusions.} We have introduced the concept of a thermal transistor based on the nanostructured coupling of the conductor to the base terminal (which is otherwise thermally isolated). The control of mesosocopic fluctuations allows for huge amplification factors. Either all-thermal transistors or non-invasive thermometers can be operated by this mechanism, depending on which current (heat or charge) is measured in the collector. Non thermal fluctuations in the gate system allow for fast switching of the gate temperature at arbitrarily low energy cost. Considering the gate as a separate system is also beneficial in opening the way to gating at a distance and reducing heat leaking. The simplicity of our idea makes it easily exportable to different kinds of systems and interactions (e.g. spin fluctuations~\cite{sothmann_magnon_2012}). We have particularised the electrostatic interaction of single-electron devices in the sequential tunneling regime, which has recently been demonstrated experimentally~\cite{holger,thgating,singh_distribution_2016,koski_on-chip_2015}. 
%The modulation of larger currents in stronger coupling regimens might be limited by higher-order processes. 
%Our results, based on a sequential tunneling approximation are limited by leakage heat induced by higher order processes~\cite{bischoff_measurement_2015,keller_cotunneling_2016,cotunneling}.
Large amplification is  found for configurations where undesired higher-order processes~\cite{bischoff_measurement_2015,keller_cotunneling_2016,cotunneling} are marginal, suggesting a way for the modulation of larger currents in stronger coupling regimes .

%{\it Acknowledgments---} 
We acknowledge financial support from the Spanish Ministerio de Econom\'ia y Competitividad via Grants No. MAT2014-58241-P and No. FIS2015-74472-JIN (AEI/FEDER/UE), and the European Research Council Advanced Grant No. 339306 (METIQUM). We also thank the COST Action MP1209 ``Thermodynamics in the quantum regime". 

%%%%%%%%%%%%%%%%%%%%%%%%%%%%%%%%

%%%%%%%%%%%%%%%%%%%%%%%%%%%%%%%%

\end{document}